\def \malpunkt{\cdot} 
\def \zl #1 #2;{\ensuremath{#1 \thinspace \textrm{#2}}}
\def \zle #1 #2 #3;{\def \test{#3}\ensuremath{\if #11\else #1 \malpunkt \fi 10^{#2} \ifx \test\empty \else \thinspace \textrm{#3} \fi}}
\def \zlr #1 #2 #3;{\ensuremath{(#1 {\thinspace\pm\thinspace} #2) \thinspace \textrm{#3}}}
\def \zler #1 #2 #3 #4;{\def \test{#4}\ensuremath{(#1 {\thinspace \pm \thinspace} #3) \malpunkt 10^{#2} \ifx \test\empty \else \thinspace \textrm{#4} \fi}}
\documentclass[aps,prl,preprint,superscriptaddress]{revtex4}

\usepackage{graphicx}
\usepackage{dcolumn}
\usepackage{bm}
\usepackage{color}

\begin{document}


\title{Frequency domain studies of current-induced magnetization dynamics in single magnetic-layer nanopillars}

\author{N. M\"{u}sgens}
\author{S. Fahrendorf}
\author{T. Maassen}\thanks {present address: Zernike Institute for Advanced Materials,
University of Groningen} \affiliation{II. Institute of Physics, RWTH
Aachen University, 52056 Aachen, Germany} \affiliation{JARA:
Fundamentals of Future Information Technology, 52074 Aachen,
Germany}

\author{A. Heiss}
\author{J. Mayer}
\affiliation{Central Facility for Electron Microscopy, RWTH Aachen
University, 52056 Aachen, Germany} \affiliation{JARA: Fundamentals
of Future Information Technology, 52074 Aachen, Germany}

\author{B. \"{O}zyilmaz}
\thanks {present address: Department of Physics, National University of
Singapore} \affiliation{II. Institute of Physics, RWTH Aachen
University, 52056 Aachen, Germany} \affiliation{JARA: Fundamentals
of Future Information Technology, 52074 Aachen, Germany}

\author{B. Beschoten}
\email{beschoten@physik.rwth-aachen.de}
\author{G. G\"{u}ntherodt}
\affiliation{II. Institute of Physics, RWTH Aachen University, 52056
Aachen, Germany} \affiliation{JARA: Fundamentals of Future
Information Technology, 52074 Aachen, Germany}

\date{\today}

\begin{abstract}
Spin transfer torque-induced high-frequency dynamics of single thin
cobalt-layer nanopillars of circular and elliptical shape have been
observed directly. Two types of precessional modes can be identified
as a function of magnetic field perpendicular to the layer
plane, excited for negative current polarity only.
They are assigned to vortex-core and transverse spin-wave excitations, which corroborate
recent model predictions.
The observed narrow linewidth of 4 MHz at room temperature indicates the high
coherence of the magnetic excitations.
\end{abstract}

\maketitle

Angular momentum conservation of the electron spin is in essence the
basis of spintronics. The spin angular momentum can be transferred
onto the magnetization of a ferromagnet (FM), thus giving rise to
so-called spin transfer torque (STT). This has been predicted
theoretically and explored experimentally for magnetic
nanostructures
\cite{JoMaMM159_Slonczewski1996_Current-DrivenExcitationofMagneticMultilayers,
PRB54_Berger1996_EmissionofSpinWavesbyaMagneticMultilayerTraversedbyaCurrent,
PRL80_Tsoi1998_ExcitationofaMagneticMultilayerbyanElectricCurrent,
PRL84_Katine2000_Current-DrivenMagnetizationReversalandSpin-WaveExcitationsinCoCuCoPillars,
PRB66_Stiles2002_AnatomyofSpin-transferTorque,
N425_Kiselev2003_MicrowaveOscillationsofaNanomagnetDrivenbyaSpin-PolarizedCurrent,
NM3_Lee2004_ExcitationsofIncoherentSpin-WavesDuetoSpin-TransferTorque,
S307_Krivorotov2005_Time-DomainMeasurementsofNanomagnetDynamicsDrivenbySpin-TransferTorques,
Kaka_Nature} to switch their magnetization, triggering also
magnetization dynamics concomitant with emission of microwave
radiation. Mainly double magnetic layer nanostructures (FM/NM/FM) of
an FM layer polarizing the electron current and an STT switching
layer with an intervening nonmagnetic metal (NM) spacer layer have
been
investigated, also regarding spin torque oscillators \cite{Kaka_Nature,NP3_Pribiag2007_MagneticVortexOscillatorDrivenbyDCSpin-polarizedC%
urrent}. However, recent experimental observations
\cite{PRL90_Ji2003_Current-InducedSpin-WaveExcitationsinaSingleFerromagneticLayer,
PRL93_Ozyilmaz2004_Current-InducedExcitationsinSingleCobaltFerromagneticLayerNanopillars,
APL88_Ozyilmaz2006_Current-InducedSwitchinginSingleFerromageneticLayerNanopillarJunctions}
and theoretical studies
\cite{PRL92_Polianski2004_Current-InducedTransverseSpin-WaveInstabilityinaThinNanomagnet,
PRB69_Stiles2004_PhenomenologicalTheoryofCurrent-InducedMagnetizationPrecession,
PRB73_Adam2006_Current-InducedTransverseSpin-WaveInstabilityinThinFerromagnetsBeyondLinearStabilityAnalysis,
PRB77_Hoefer2008_ModelforaCollimatedSpin-WaveBeamGeneratedbyaSingle-LayerSpinTorqueNanocontact}
suggest that even an unpolarized current passing perpendicular to a
single thin FM layer can cause an STT on this FM layer. Spin
filtering effects by a single FM layer contacted by NM leads creates
spin accumulation of opposite sign at both interfaces. The
associated torques of opposite sign may cancel each other.
Asymmetric leads, however, break the symmetry of the single magnetic
layer (NM/FM/NM) structure and yield a net STT on the magnetization
for one current polarity only
\cite{PRL92_Polianski2004_Current-InducedTransverseSpin-WaveInstabilityinaThinNanomagnet}.
Requirement is that the spins diffuse in the NM leads along the
interfaces and that the FM magnetization is nonuniform or
inhomogeneous. This mechanism was not included in the Slonczewski
model
\cite{JoMaMM159_Slonczewski1996_Current-DrivenExcitationofMagneticMultilayers}.
The direction of the torque is to align the local magnetization with
the direction of the accumulated spins.

For single, extended FM layers studied by means of point contacts
the predicted spin-torque induced excitations depend on competing
Oersted and external magnetic fields
\cite{PRB77_Hoefer2008_ModelforaCollimatedSpin-WaveBeamGeneratedbyaSingle-LayerSpinTorqueNanocontact}.
For thick FM layers the formation of a nanodomain underneath the
nanocontact obtains a more intricate experiment
\cite{PRL90_Ji2003_Current-InducedSpin-WaveExcitationsinaSingleFerromagneticLayer,PRL93_Chen2004_Current-DrivenSwitchinginaSingleExchange-BiasedFerr%
omagneticLayer}. For single FM layers laterally confined to a
nanopillar, current-induced excitations of spin waves transverse to
the current flow direction have been predicted
\cite{PRL92_Polianski2004_Current-InducedTransverseSpin-WaveInstabilityinaThinNanomagnet,
PRB73_Adam2006_Current-InducedTransverseSpin-WaveInstabilityinThinFerromagnetsBeyondLinearStabilityAnalysis}.
Assuming additionally a nonuniform magnetization normal to the layer
plane for the net spin torque a lowest mode is found, which is
nonuniform both parallel and perpendicular to the layer plane
\cite{PRB69_Stiles2004_PhenomenologicalTheoryofCurrent-InducedMagnetizationPrecession,
JoAP97_McMichael2005_MagneticNormalModesofNanoelements}. First
experimental observations of current induced resistance changes of
single nanosized FM layers have been reported
\cite{PRL93_Ozyilmaz2004_Current-InducedExcitationsinSingleCobaltFerromagneticLayerNanopillars,
APL88_Ozyilmaz2006_Current-InducedSwitchinginSingleFerromageneticLayerNanopillarJunctions}.
Only by analogy to magnetic double layer structures have they been
assigned to the onset of possible magnetic excitations. However, a
direct experimental proof of the high frequency magnetization
dynamics and identification of the excited magnetic modes is still
missing.

Using frequency domain spectroscopy, we observe for a single
nano-confined thin FM layer directly two types of characteristic
eigenmodes triggered by a dc current perpendicular to the layer
plane. These modes are assigned to vortex-core and transverse
spin-wave excitations deduced from their characteristic frequency vs
current or vs field dependences. This is in contrast to previous
static studies of differential-resistance $dV/dI$ vs current $I$,
which help only in identifying mode onsets. Asymmetries are observed
with respect to current polarity for spin excitations and field
polarity for microwave emission power as well as mode overtones. The
small linewidth of only 4 MHz for Co at room temperature (RT) proves
the high coherence due to nonlocal STT effects. Our results confirm
recent qualitative predictions of current-induced spin-wave
excitations in single FM layers. Moreover, they pave the way for
technological applications due to relaxed constraints on
magnetization uniformity and elimination of retroactions in dynamics
between switching and polarizing, fixed magnetic layers
\cite{N425_Kiselev2003_MicrowaveOscillationsofaNanomagnetDrivenbyaSpin-PolarizedCurrent}.

The samples have been fabricated using a focused ion-beam assisted
nanostencil mask technique
\cite{JoAP101_Ozyilmaz2007_Focused-Ion-BeamMillingBasedNanostencilMaskFabricationforSpinTransferTorqueStudies}.
The thin layer stack of 10~nm~Cu/15~nm~Co/150~nm~Cu (Fig. 1(a)) is
deposited by means of molecular beam epitaxy in prefabricated
undercut templates on top of a Pt bottom electrode. The latter acts
additionally as a spin sink ensuring the asymmetry of the electrical
leads. The investigated nanopillars have either circular (diameter
$d\approx 40$~nm) or elliptical ($50\times 100$~nm$^2$)
cross-sections. The device is connected to microwave probes and a dc
current flows through a bias-tee, along with a $100~\mu$A ac
current. The high frequency components of the resistance changes due
to magnetization dynamics are analyzed using a spectrum analyzer
with a bandwidth of the circuit of $0.1\leq f \leq 20$~GHz. The
electron flow from the thick to the thin Cu layer lead, known to
cause a destabilizing or switching torque, is defined as the
negative current polarity. All measurements are performed at RT and
for each power spectrum $\Delta P(I,H)$ presented the current
independent background noise has been subtracted.

\begin{figure}
\includegraphics{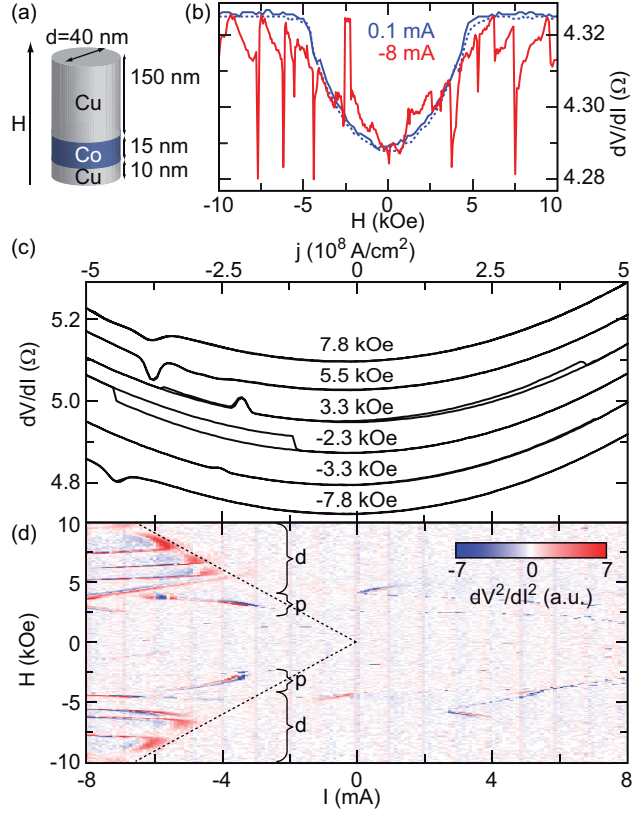}
\caption{\label{fig1} (color online) (a) Schematic of a circular
single Co-layer nanopillar with asymmetric Cu leads. (b) $dV/dI$ vs
$H$ at $I= 0.1$~mA (backward field-sweep direction is marked by
dotted line) and at $-8$~mA (the increased resistance due to the
increased current has been subtracted). (c) $dV/dI$ vs $I$ for
representative $H$ values. (d) False color plot of $d^2V/dI^2$ vs
$I$ for $|H|\leq 10$~kOe. The black dashed line indicates the
boundary of the resistance features; $p$ and $d$ denote the peak and
dip regimes in $dV/dI$ in (c), respectively.}
\end{figure}

The anisotropic magnetoresistance (AMR) of a single magnetic layer
nanopillar at almost zero dc bias (0.1~mA) and a magnetic field
oriented perpendicular to the sample plane (ptp) is shown in Fig.
1(b). Starting from $H=0$ in a low resistance state $dV/dI$
saturates in a high resistance state at magnetic fields above 5~kOe,
indicating a ptp magnetization alignment, parallel to $H$. This
field is reduced from the out-of-plane demagnetization field of Co
for the unpatterned film, because of the Co layer confined in the
nanopillar yielding an out-of-plane demagnetization factor of 6.72
in cgs units (compared to $4\pi$ of the unpatternded film). This was
calculated from 6-fold integration of dipole-dipole interaction for
the given geometry \cite{Lee-private-communication}. For high
negative current values of $I = -8$~mA ($j
\approx-5\times10^8$~A/cm$^2$) there are resistance anomalies
superimposed on this AMR curve in the form of pronounced dips and
peaks, which are absent for positive current values (not shown).
Typical $dV/dI$ changes as a function of $I$ are shown in Fig. 1 (c)
for different $H$ of both polarities. To better visualize the
current and field dependence of these resistance curves and for a
better distinction from the parabolic background (Fig. 1(b,c)) we
plot in Fig. 1(d) the second derivative $d^2V/dI^2$ (color coded) as
a function of $I$ and applied ptp-$H$-field. For $I>0$ no pronounced
dips and peaks are observed, consistent with predictions
\cite{PRL92_Polianski2004_Current-InducedTransverseSpin-WaveInstabilityinaThinNanomagnet}.
For $I<0$ and increasing absolute field values we first observe
peaks ($p$) in $dV/dI$, which represent an increase in junction
resistance. These features shift with increasing fields to higher
negative current values. The opposite dependence is observed for the
dips ($d$), appearing at slightly higher absolute magnetic fields.
The dips indicate a decrease in junction resistance, which has been
theoretically attributed to spin-wave excitations
\cite{PRL92_Polianski2004_Current-InducedTransverseSpin-WaveInstabilityinaThinNanomagnet,
PRB73_Adam2006_Current-InducedTransverseSpin-WaveInstabilityinThinFerromagnetsBeyondLinearStabilityAnalysis}.
As shown in Fig 1(c) there exists an asymmetry in the amplitude and
current value of peaks and dips with regard to the field polarity.
Hysteretic switching effects is observed in a small range near $H=-2.3$~kOe.
Similar results are present in elliptically shaped
nanopillars (see Fig. 3(d)), consistent with recent experimental
work
\cite{PRL93_Ozyilmaz2004_Current-InducedExcitationsinSingleCobaltFerromagneticLayerNanopillars,
APL88_Ozyilmaz2006_Current-InducedSwitchinginSingleFerromageneticLayerNanopillarJunctions}.

\begin{figure}
\includegraphics{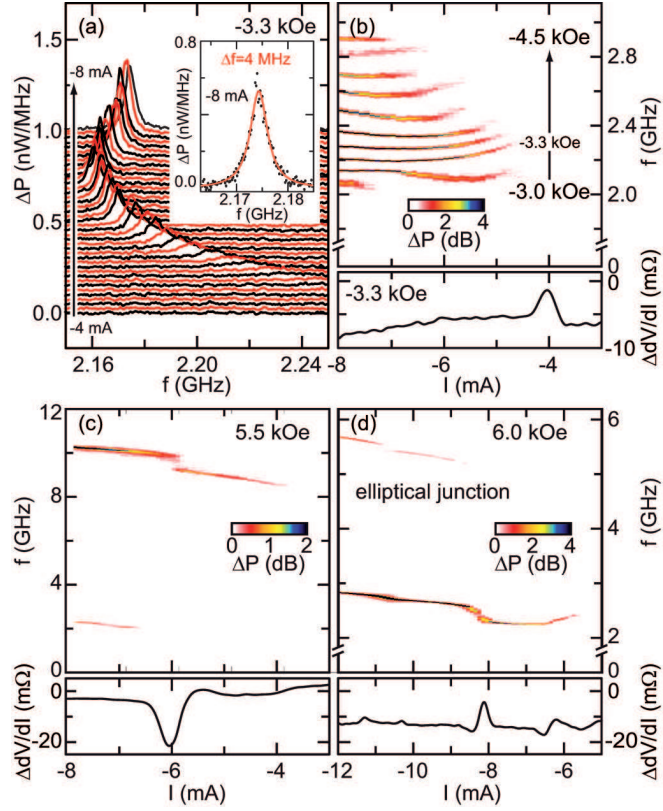}
\caption{\label{fig2} (color online). Microwave power spectra
$\Delta P(I,H,f)$ for circular (a)-(c) and elliptical (d) nanopillar
at RT. (a) $\Delta P(f)$ for $H = -3.3$~kOe and $- 8$~mA $\leq I
\leq -4$~mA (curves are shifted for clarity); inset: Lorentzian fit
to the spectra taken at $I = -8$~mA. (b) Current dependence of
$\Delta P$ in logarithmic scale
at different field values ($H
=-3.0$,$-3.1$,$-3.3$,$-3.5$,$-3.7$,$-3.9$,$-4.0$,$-4.2$,$-4.3$,$-4.5$~kOe);
bottom: $\Delta dV/dI$ vs $I$ at $H = -3.3$~kOe. (c)
$\Delta P$ vs $I$
at $H = 5.5$~kOe;
bottom: $\Delta dV/dI$ (parabolic background subtracted) vs $I$ at
$H = 5.5$~kOe. (d) $\Delta P$
vs $I$
at $H = 6.0$~kOe; bottom: $\Delta dV/dI$ vs $I$ at 6~kOe.}
\end{figure}

First we consider the frequency spectrum of the circular sample for
$H=-3.3$~kOe and currents from $-4$~mA to $-8$~mA (Fig. 2(a)). At
$I=-4.8$~mA, slightly beyond the peak observed in $\Delta dV/dI$
(Fig. 2(b), bottom), we begin to resolve microwave emission at
$f\approx 2.23$~GHz. $\Delta dV/dI$ denotes that the parabolic
background has been subtracted. As the negative current is increased
in magnitude the initial signal grows in amplitude and shifts to
smaller frequencies until $I\approx -6.8$~mA, beyond which the
excitation frequency increases again. The very narrow linewidth of
4~MHz for Co at RT (inset Fig. 2(a)) demonstrates the unexpected
high coherence of the observed mode. As one can see in Fig. 2(b) the
above described "low" frequency ("low-f") mode is representative for
the peak regime (compare Fig. 1(d)). The onset of microwave emission
in Fig. 2(b) shifts to higher negative current values for increasing
negative fields consistent with the shift of the peak position in
the $d^2V/dI^2$ vs $I$ diagram (Fig. 1(d)).
Please note that the output power $\Delta P$ does not directly scale
with the peak amplitude in the $dV/dI$ vs $I$ curves (Fig. 1(c)),
since we observe lower microwave power at $H>0$ (not shown) where
the peak amplitudes are larger.

Besides the "low-$f$" mode an additional high frequency mode is
observed (Fig. 2(c)). The onset of the "high-$f$" mode is related to
a decrease in $dV/dI$ vs $I$ near -4~mA (bottom panel of Fig. 2(c))
and does not correspond to a higher harmonics of the "low-$f$" mode.
The frequency of this high-$f$ mode with a typical linewidth of
$\sim 10$~MHz near $I=-5$~mA increases linearly with increasing
negative current. The microwave signal exhibits a frequency jump
near $-6$~mA linked to a strong dip in the differential resistance.
The jump is attributed to transitions between different excited
spin-wave modes as theoretically predicted
\cite{PRB69_Stiles2004_PhenomenologicalTheoryofCurrent-InducedMagnetizationPrecession}.
The bistability of both low-$f$ and high-$f$ mode vanishes for
further increasing fields (Fig. 3(a)). In comparison to single mode
excitations the modes in the bistability regime have broader
linewidths of 10~MHz and 90~MHz for the low-$f$ and high-$f$ mode,
respectively, at $I=-8$~mA and $H=5.5$~kOe. This indicates a
decrease in temporal coherence
\cite{PRB76_Krivorotov2007_Large-amplitudeCoherentSpinWavesExcitedbySpin-polarizedCurrentinNanoscaleSpinValves}.
$\Delta dV/dI$ of nanopillars with elliptical cross-section in Fig.
2(d), bottom, shows alternating dips and peaks. Surprisingly, only
one narrow mode together with its first harmonic is observable. The
assignment to peaks and dips is not as straightforward and more
complex.

\begin{figure}
\includegraphics{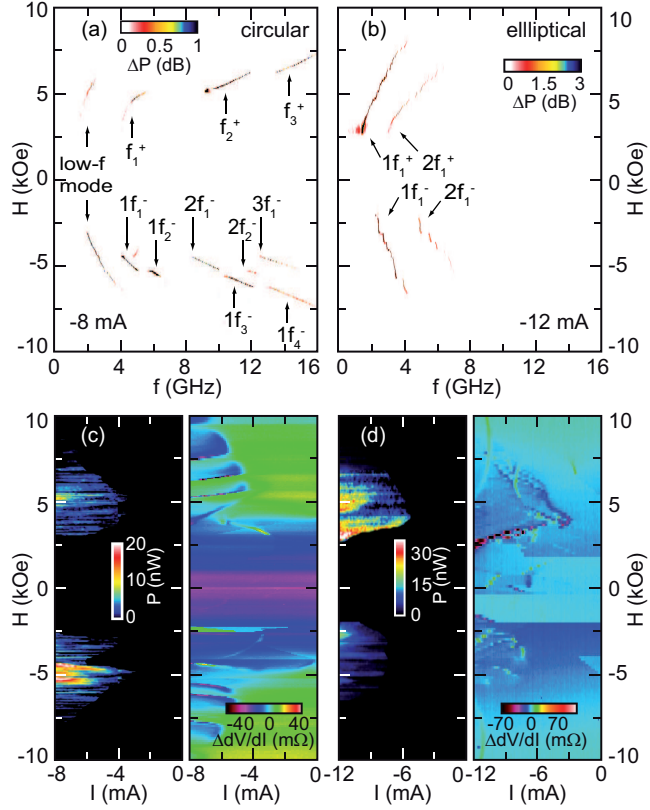}
\caption{\label{fig3} (color online). (a) Emitted microwave power
$\Delta P$
of the circular nanopillar as a function of out-of-plane magnetic field at
$I=-8$~mA. (b) Field dependence of $\Delta P$
of the elliptical nanopillar at $I=-12$~mA. (c) Circular nanopillar,
left panel: Integrated microwave power in false color plot vs $I$
and $H$; right panel: false color plot of the differential
resistance $\Delta dV/dI$ (parabolic background subtracted) vs $I$
and $H$. (d) Same as in (c) for the elliptical nanopillar.}
\end{figure}

To further explore the diversity of the observed modes we discuss
the field dependence of the emitted microwave power for $I<0$.
For the circular sample (Fig. 3(a)) multiple mode frequencies are
observed for both field polarities. Besides the low-$f$ mode these
are the high-$f$ modes $f_i^+$, $i=1,2,3,\ldots$, for $H>0$ and
$f_i^-$ for $H<0$ together with their higher harmonics $n\cdot
f_i^-$, $n=1,2,3,\ldots$. For a given mode the frequency corresponds
to an eigenmode of the Co nanodisc and increases nearly linearly
with the applied field as expected from the theoretical models
\cite{PRB73_Adam2006_Current-InducedTransverseSpin-WaveInstabilityinThinFerromagnetsBeyondLinearStabilityAnalysis,
JoAP97_McMichael2005_MagneticNormalModesofNanoelements}. A field
asymmetry concerns the measured frequencies $f^+_i$ and $f^-_i$,
which show also jumps.

For the elliptically shaped nanopillar the field dependence of the
mode frequencies is less complex, because only one mode together
with its first harmonic is excited (Fig. 3(b)). Again a strong
asymmetry appears concerning the field polarity, since $n\cdot
f^+(H) \neq n\cdot f^-(H)$ and frequency jumps appear for negative
fields only. Hence the asymmetry in the transport measurements
($\Delta dV/dI$ vs $I$ in Fig. 3(d), right panel) reflects that of
the observed frequency behavior as well as that of the integrated
microwave power (Fig. 3 (d), left panel). The correlation of the
resistance changes (Fig. 3 (c,d), right panels) to the microwave
power output (Fig. 3 (c,d), left panels) is clearly visible, but
there is still a lack in understanding the link to the quantitative
magnitude of the power.

Micromagnetic calculations of excited normal modes for circular and
elliptical nanoelements exhibit low frequency uniform precession
modes and higher frequency modes of different (even/odd) symmetries
\cite{JoAP97_McMichael2005_MagneticNormalModesofNanoelements}. For
elliptical shapes the many symmetry modes are merging into fewer
ones as calculated from spin-wave dispersion for discrete wave
vectors. This may explain the fewer high frequency modes of
elliptical compared to circular elements (Fig. 3a,b), besides the
lower elliptical symmetry. A more rigorous treatment has to analyze
for the lowest critical current of modes being driven unstable.
Quantitative predictions have failed so far because of the
difficulty in combining spin diffusion with micromagnetic
simulations. It has been argued that inhomogeneities in the
magnetization transverse to the current flow direction reduce the
resistance of a single magnetic layer due to a mixing of spin
channels. As the high-$f$ mode of circular pillars is linked to the
dips in the differential resistance one could attribute this mode to
transverse spin waves, which are expected in single magnetic layer
nanopillar devices
\cite{PRL92_Polianski2004_Current-InducedTransverseSpin-WaveInstabilityinaThinNanomagnet,
PRB73_Adam2006_Current-InducedTransverseSpin-WaveInstabilityinThinFerromagnetsBeyondLinearStabilityAnalysis}.
In order to compare our measurements for the circular nanopillar
with calculated eigenfrequencies of a circular Co nanoelement
\cite{JoAP97_McMichael2005_MagneticNormalModesofNanoelements}, we
extrapolate the high frequency $f_3^+$ mode (Fig. 3(a)) to higher
fields. The extrapolated value of about 40~GHz at 20~kOe agrees well
with the calculated value for transverse spin waves.

The peaks in $dV/dI$, equivalent to an increase in junction
resistance, may be explained by inhomogeneities in the magnetization
normal to the thin film plane, i.e. due to a GMR-like effect. These
peaks are most pronounced at relatively low fields ($H\sim 3-5$~kOe)
for the circular pillar (Fig. 2(b)). A circular pillar with diameter
of d $\approx40$~nm and a Co thickness of 15~nm, together with the
current induced Oersted field and the external ptp-H field, enables
the formation of a vortex inside the FM layer
\cite{S298_Wachowiak2002_DirectObservationofInternalSpinStructureofMagneticVortexCores,
S289_Shinjo2000_MagneticVortexCoreObservationinCircularDotsofPermalloy,
PRL94_Ding2005_MagneticBistabilityofCoNanodots,
PRB73_Urazhdin2006_EffectsofCurrentontheMagneticStatesofPermalloyNanodiscs}.
The lowest excitation mode for a vortex is the gyroscopic precession
of the core with frequencies of the order of f $\sim 1$~GHz
\cite{S304_Choe2004_VortexCore-DrivenMagnetizationDynamics},
comparable to the mode frequency in Fig. 2(b). It has been shown
\cite{NP3_Pribiag2007_MagneticVortexOscillatorDrivenbyDCSpin-polarizedCurrent}
that the vortex core dynamics can be excited by spin-polarized
currents. In that case micromagnetic simulations demonstrate a
vortex-core precession with unequal trajectories on either side of
the FM layer, resulting from inhomogeneities in the magnetization
perpendicular to plane. Therefore we suppose the low-$f$ mode to be
a vortex-core like precession. The more complex resistance changes,
observed in Fig. 2(d), bottom, may give evidence that the mode near
3~GHz and its overtone originate from multivortex-state precessions
\cite{JoAP95_Okuno2004_TwoTypesofMagneticVortexCoresinEllipticalPermalloyDots}.

In conclusion, we have observed current-induced triggering of highly
coherent spin waves in single Co layer nanopillars at RT by an
unpolarized current for out-of-plane magnetic fields. Dynamical
modes at different frequencies are found and a possible correlation
to transverse spin waves and vortex-core dynamics is discussed for
circular and elliptical cross-sections of the nanopillars. These
experimental data require more detailed simulations which we hope to
have stimulated by this work.

We acknowledge useful discussion with K.J. Lee. This work was
supported by DFG through SPP 1133.
%

\end{document}